\documentclass{article}
\usepackage{spconf}
\usepackage[utf8]{inputenc} 
\usepackage[T1]{fontenc}
\usepackage{enumitem,amsmath,graphicx}
\usepackage{amsfonts}             
\usepackage{microtype}
\usepackage{nicefrac}    
\usepackage{color, colortbl}
\usepackage{xcolor}
\usepackage{booktabs}
\usepackage{mathrsfs}

\usepackage{setspace}
\usepackage{array}
\usepackage{threeparttable}
\usepackage{subcaption}
\usepackage[thinlines]{easytable}
\usepackage{breqn}
\usepackage{multirow}

\title{Expression of White and Gray Matter Autism Functional Network Connectivity Signatures in Infancy}
\name{Qiang Li$^{1}$, Jingyu Liu$^{1,2}$, Zening Fu$^{1}$, Armin Iraji$^{1,2}$, Sarah Shultz$^{3,4}$, Vince D. Calhoun$^{1,2}$}
\address{$^{1}$Tri-Institutional Center for Translational Research in Neuroimaging and Data Science (TReNDS), \\
Georgia State, Georgia Tech, and Emory University, Atlanta, GA, United States \\
$^{2}$Department of Computer Science, Georgia State University, Atlanta, GA, United States\\
$^{3}$Marcus Autism Center, Children's Healthcare of Atlanta, Atlanta, GA, United States\\
$^{4}$Division of Autism \& Related Disabilities, Department of Pediatrics, \\ Emory University School of Medicine, Atlanta, GA, United States}

\begin{document}
\maketitle
\begin{abstract}
Functional connectivity studies have provided valuable insights into altered brain organization in autism spectrum disorder (ASD). However, most functional connectivity studies focus exclusively on either gray matter (GM) regions or white matter (WM) tracts and examine specific age groups, rather than investigating their functional interactions across brain development. This limits our understanding of how GM-WM functional interactions evolve developmentally and may obscure important aspects of brain network organization in ASD. We combined integrated WM+GM functional network connectivity (FNC) with brain-wide risk scores (BRSs) to quantify ASD-associated pattern expression in ABIDE I/II and an independent infant cohort. We employed a NeuroMark framework with predefined WM and GM intrinsic connectivity networks templates, to estimate subject-level FNC. First, we identified two distinct patterns of altered WM+GM FNC in ASD: GM-related hypoconnectivity centered on the paralimbic network and WM-GM hyperconnectivity centered on the frontal network. These findings highlight the involvement of higher-order cognitive networks and the importance of WM-GM functional interactions in characterizing ASD-related connectivity alterations. Second, infants with a higher familial likelihood of autism exhibited greater mean expression of ASD-associated FNC patterns than those with a lower familial likelihood, as quantified by BRS. These group-level findings do not establish developmental persistence or predict diagnostic outcomes at the individual level\footnote{\textbf{2027 IEEE 24th International Symposium on Biomedical Imaging}}.
\end{abstract}


\section{Introduction}
\label{sec:intro}
Functional connectivity has been extensively investigated in ASD, providing promising insights into altered brain network organization and helping identify neuroimaging patterns associated with the condition~\cite{hull2017resting}. These approaches have also facilitated cross-species comparisons, offering opportunities to investigate the conservation of neural connectivity patterns across species~\cite{pagani2026autism,LI2026122241}. However, most existing studies have examined GM or WM regional functional connectivity separately within specific age groups, potentially overlooking interactions between these two tissue types across brain development. Considering GM and WM functional connectivity separately may therefore provide an incomplete picture of brain network organization in ASD across brain development, as their interactions may reveal additional patterns of atypical functional connectivity that are not detectable when either tissue type is examined in isolation. 
 
To address this gap, we combined WM+GM functional network connectivity (FNC, i.e. connectivity between whole brain networks) with BRS to quantify the expression of ASD-associated patterns in infancy. Using data from the Autism Brain Imaging Data Exchange (ABIDE) I and II datasets~\cite{di2014autism} and a 0\mbox{-}6-month-old infant dataset~\cite{li2026deciphering}, we investigated FNC patterns across WM and GM and identified distinct ASD-related alterations across development. Furthermore, we identified two distinct patterns of ASD-associated WM+GM functional connectivity, including hypoconnectivity involving GM and hyperconnectivity between WM and GM in the ABIDE cohort. BRS quantified the expression of ABIDE-derived ASD connectivity signatures in infants. These results highlight the importance of considering WM-GM functional interactions to obtain a more comprehensive understanding of altered functional network organization and its heterogeneity in ASD. 
\begin{table*}[!ht]
\centering
\caption{\textbf{Demographic characteristics of the ABIDE-I, ABIDE-II, and infant cohort.}}
\label{tab:demographics}
\renewcommand{\arraystretch}{1.1}
\setlength{\tabcolsep}{6pt}
\begin{threeparttable}
\resizebox{\textwidth}{!}{
\begin{tabular}{lccc|ccc|cc}
\toprule
& \multicolumn{3}{c|}{\textbf{ABIDE-I}}
& \multicolumn{3}{c|}{\textbf{ABIDE-II}}
& \multicolumn{2}{c}{\textbf{Infant Cohorts}} \\
\cmidrule(lr){2-4}
\cmidrule(lr){5-7}
\cmidrule(lr){8-9}
\textbf{Demographic}
& \textbf{CN}
& \textbf{ASD}
& \textbf{$p$-value}
& \textbf{CN}
& \textbf{ASD}
& \textbf{$p$-value}
& \textbf{LL infants}
& \textbf{HL infants} \\
\midrule
\textbf{Subjects}
& 246
& 215
& --
& 592
& 521
& --
& 84/ 203[scans]
& 59/ 127[scans] \\
\textbf{Age}
& $17.20 \pm 8.21$ years
& $18.29 \pm 9.39$ years
& 0.1859
& $14.97 \pm 9.31$ years
& $14.75 \pm 9.00$ years
& 0.6858
& $103.13 \pm 50.36$ days
& $121.98 \pm 40.52$ days \\
\textbf{Corrected age}
& --
& --
& --
& --
& --
& --
& $99.10 \pm 51.69$ days
& $118.96 \pm 41.24$ days \\
\textbf{Sex}
& &
&
&
&
&
&
& \\[-4pt]
\quad Male
& 202 (82.1\%)
& 196 (91.2\%)
& \textbf{0.0048}
& 412 (69.6\%)
& 444 (85.2\%)
& \textbf{<0.0001}
& 121 [scans]
& 60 [scans]\\
\quad Female
& 44 (17.9\%)
& 19 (8.8\%)
&
& 180 (30.4\%)
& 77 (14.8\%)
&
& 82 [scans]
& 67 [scans]\\
\bottomrule
\end{tabular}
}
\begin{tablenotes}[flushleft]
    \footnotesize
    \item[] \parbox[t]{\textwidth}{%
        \textit{Note.} Values are presented as mean $\pm$ standard
        deviation for age and corrected gestational age, and as number
        (percentage) for sex in the ABIDE cohorts, and as number for
        sex in the infant cohort. Age is reported in years for the
        ABIDE cohorts and in days for the infant cohorts. Corrected
        age is applicable only to the infant cohort. $p$-values for
        age were obtained using two-sample $t$-tests. $p$-values for
        sex were obtained using Pearson's $\chi^2$ tests comparing
        CN and ASD groups within each ABIDE dataset. Bold values
        indicate statistically significant differences ($p<0.05$).}
\end{tablenotes}
\end{threeparttable}
\end{table*}

\section{Methodology}
\label{sec:meths}
\subsection{Data Information and Preprocessing}
We analyzed resting-state fMRI data from the ABIDE I and II datasets and an infant dataset (\textbf{Table~\ref{tab:demographics}}). Additional information regarding the datasets and MRI acquisition parameters is available in~\cite{di2014autism,li2026deciphering}. The study included 1,574 participants from ABIDE I and II, comprising 736 individuals with ASD and 838 typically developing controls (CN), as well as a separate dataset of 0\mbox{-}6-month-old infants, including 84 infants with lower (LL) and 59 with higher (HL) familial likelihood of autism. The imaging data were preprocessed using a standard pipeline that included rigid-body motion correction, slice-timing correction, nonlinear registration to the Montreal Neurological Institute (MNI) standard space, and spatial smoothing with a Gaussian kernel with a full width at half maximum of 6 mm. The infant dataset underwent the same preprocessing pipeline, except for the distortion correction and the normalization, where the images were first registered to a 3-month-old infant brain template~\cite{chen20224d,fu2026robustness} and then transformed into MNI space.

\subsection{Spatially Constrained Independent Component Analysis}
To estimate subject-level spatial maps and their associated time courses, we employed a NeuroMark framework~\cite{du2020neuromark}. This data-driven approach decomposes fMRI data into spatially independent components, while incorporating prior spatial information to guide the identification of intrinsic connectivity networks (ICNs). Specifically, we used the NeuroMark 2.2 template including both WM and GM networks~\cite{Iraji2022CanonicalAR,itkyal2025evidence} as a spatial constraint to guide the decomposition and obtain subject-specific spatial maps and their corresponding time courses. The template comprised 202 ICNs, including 97 WM ICNs and 105 GM ICNs (\textbf{Fig.\ref{fig:1}A}). Further details on template construction are provided in~\cite{Iraji2022CanonicalAR,itkyal2025evidence}.

\subsection{Statistical Analysis}
Following scICA, subject-level FNC was estimated by calculating pairwise Pearson correlation coefficients between the time courses of the 202 ICNs, yielding 20,301 unique connections per subject. We first harmonized the FNC measures across acquisition sites using ComBat~\cite{chen2022mitigating} to mitigate site-related effects while preserving biological variability. We then employed general linear models (GLMs) to assess group differences in ABIDE FNC between individuals with ASD and CN, with age and sex included as covariates. The resulting regression coefficients ($\beta$) were used to characterize the direction and magnitude of group differences. Multiple comparisons were corrected using the false discovery rate (FDR) procedure, with statistical significance defined as $q < 0.05$. Significant connections were classified as hypoconnectivity or hyperconnectivity in ASD relative to CN based on the direction of the estimated group effects. These significant connections were subsequently examined to characterize network-specific alterations in WM+GM functional connectivity in ASD. In addition, we estimated BRSs~\cite{yan2024brainwide} from FNC patterns in both ABIDE and infant cohorts to quantify ASD-associated pattern expression. BRS contrasts distances to ASD and CN references, indexing relative pattern similarity rather than diagnostic probability. We compared BRS distributions across groups to assess ASD-CN differences in ABIDE and HL-LL differences in infants.

\begin{figure*}
    \centering
    \begin{minipage}{0.43\textwidth}
        \centering
        \small{(A) WM+GM Templates of Spatial Maps}\\[1mm]
        \includegraphics[width=\textwidth, height=16cm]{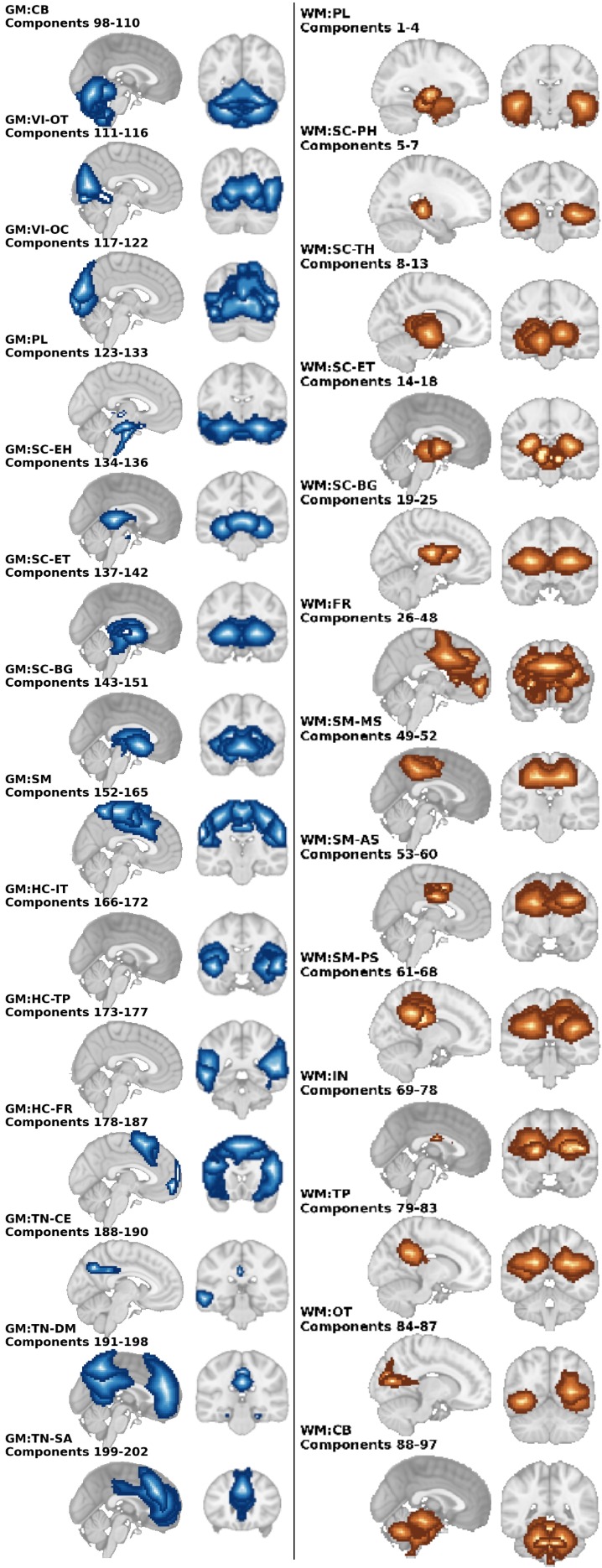}
         \small{(B) Average Harmonized CN \& ASD FNC (WM+GW)}\\[1mm]
        \includegraphics[width=1.1\textwidth, height=7.5cm]{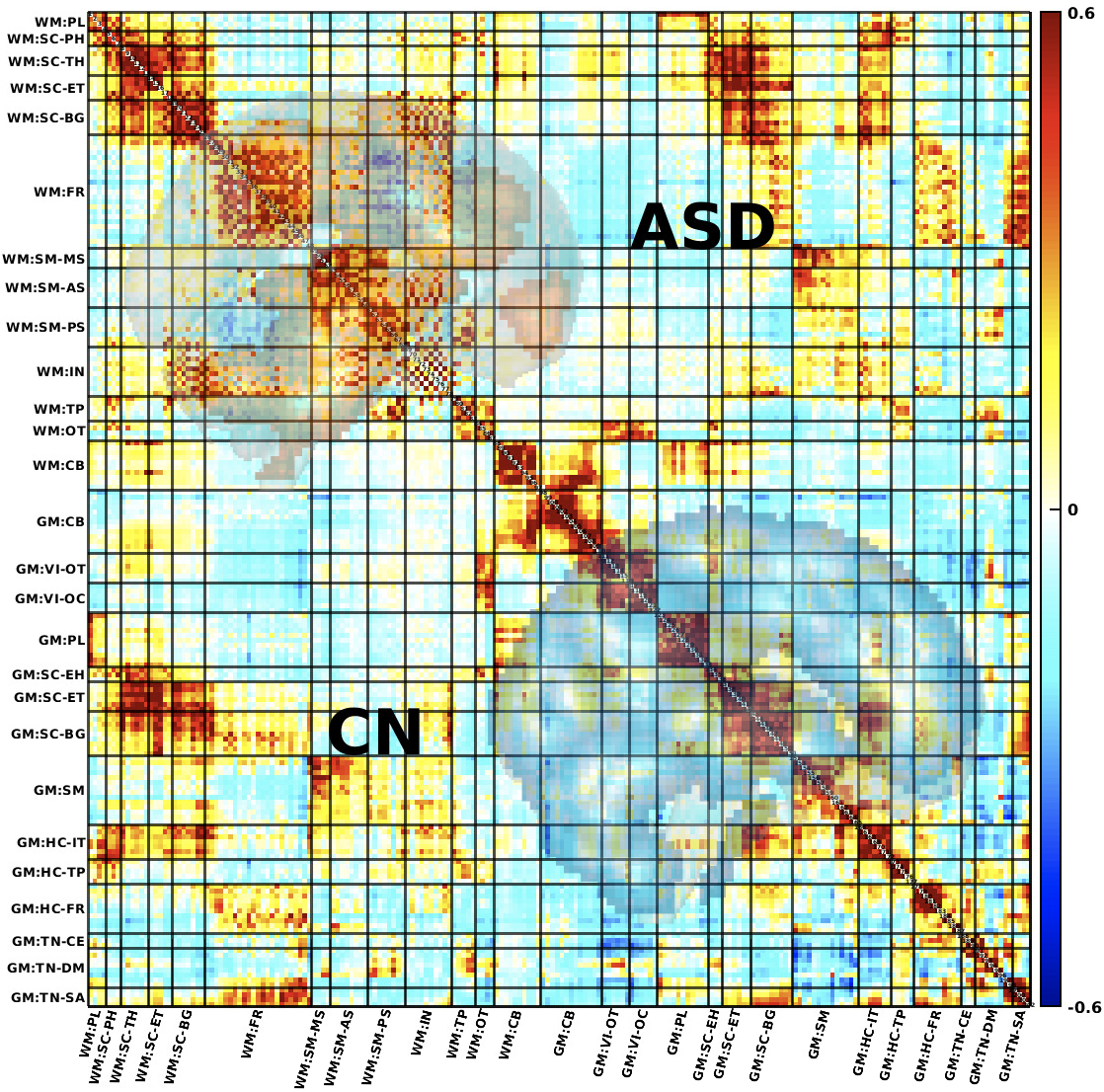}
    \end{minipage}
    \hfill
    \begin{minipage}{0.06\textwidth}
        \centering
        \small{\textbf{ABIDE}}\\[1mm]
        \includegraphics[width=\textwidth]{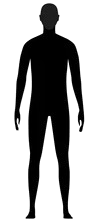}
    \end{minipage}
    \hfill
    \begin{minipage}{0.5\textwidth}
        \centering
        \small{(C) ASD vs. CN FNC: Adjusted Group Effects}\\[1mm]
        \includegraphics[width=0.97\textwidth, height=7.5cm]{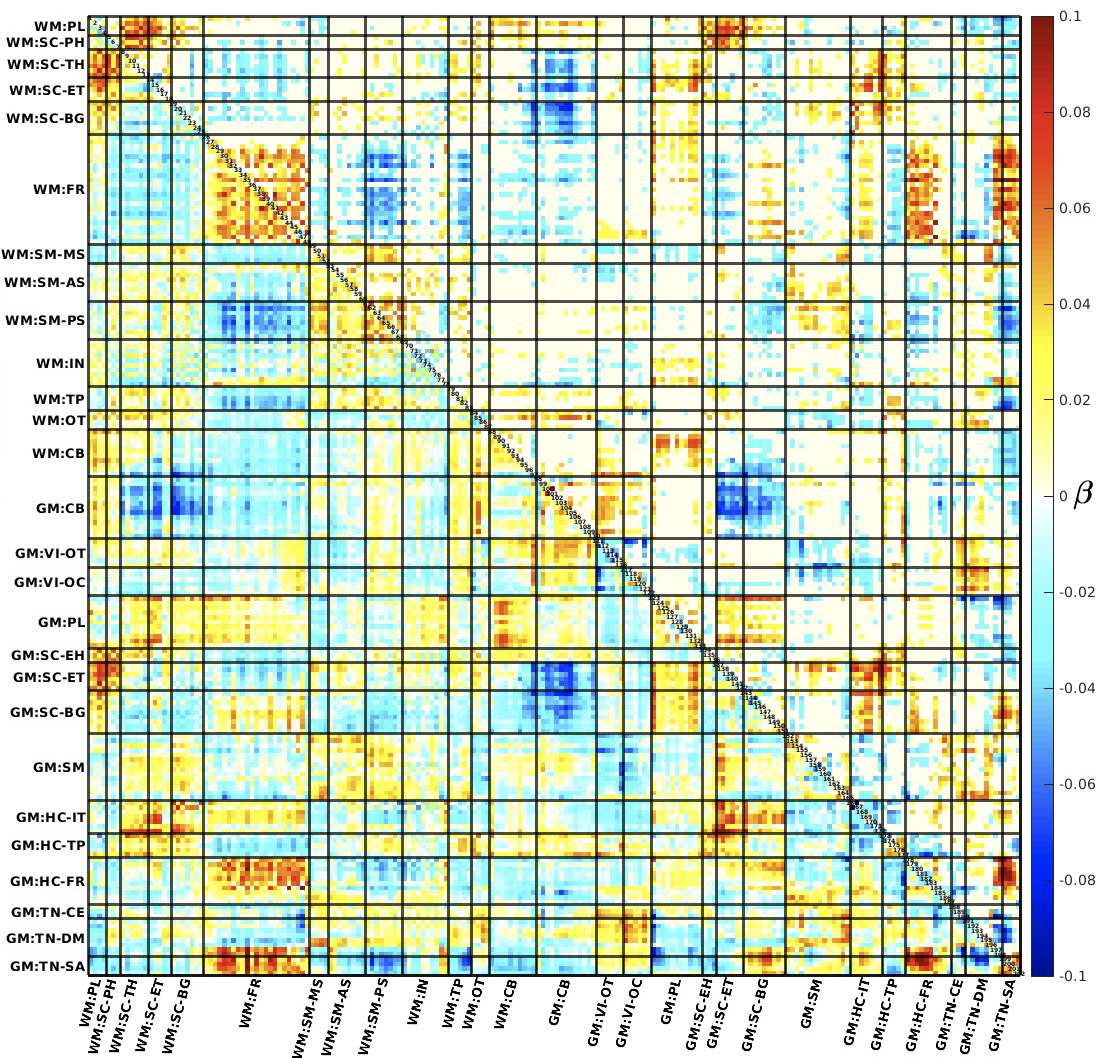}
        \small{(D) GM-GM \textcolor{blue}{Hypoconnectivity (Hub: ICN 123)} \\ WM-GM \textcolor{red}{Hyperconnectivity (Hub: ICN 35)} in ASD}\\[1mm]
        \includegraphics[width=0.86\textwidth, height=7.2cm]{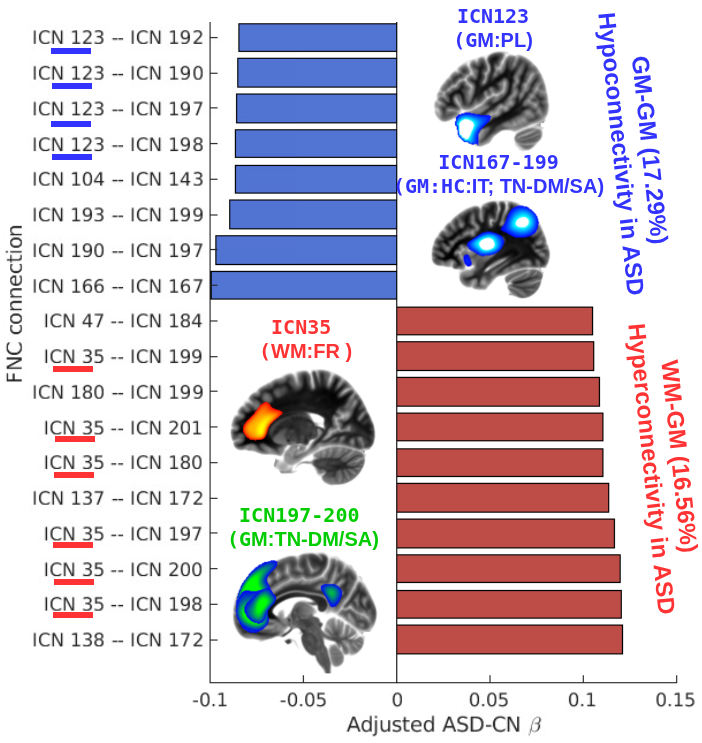} \\
        \small{(E) ASD-Associated Pattern Expression in Infancy}\\[1mm]
        \includegraphics[width=0.7\textwidth]{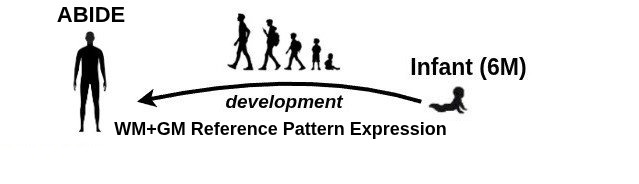}
        \includegraphics[width=0.97\textwidth, height=6.5cm]{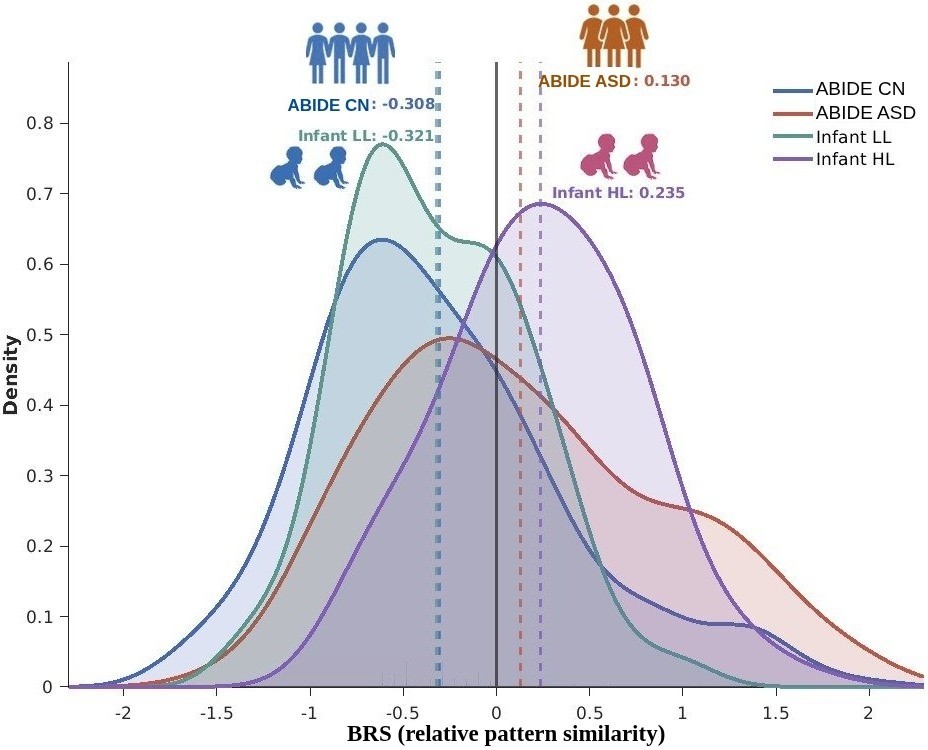}
    \end{minipage}
    \label{fig:1}
\end{figure*}
\begin{figure*}
\addtocounter{figure}{0}
\caption{\textbf{Brain connectome signatures of ASD across development.} (A) GM (ICNs:98-202)+WM (ICNs:1-97) template spatial maps of ICNs. (B) The average harmonized FNC for CN (lower triangular) and ASD (upper triangular) participants. (C) Whole-brain FNC was analyzed using a GLM, with CN as the reference group. The lower triangle shows the estimated ASD-CN $\beta$ coefficients for all connections, whereas the upper triangle displays only connections that survived FDR correction at $q<0.05$. Color represents the direction and magnitude of the group effect, with positive $\beta$ values indicating greater connectivity in ASD and negative $\beta$ values indicating lower connectivity in ASD relative to CN. (D) The top significant ($q<0.05$) hypoconnectivity and hyperconnectivity findings in ASD were selected to characterize network-specific connectivity alterations relative to CN. (E) BRS distributions for ABIDE ASD/CN and infant HL/LL groups.}
\label{fig:1}
\end{figure*}

\section{Results}
\label{sec:res}
\subsection{WM+GM FNC Differences Between CN and ASD}
The average harmonized WM+GM FNC matrices were computed separately for the CN and ASD  groups, capturing within-WM, within-GM, and cross-tissue WM-GM functional connections. (\textbf{Fig.\ref{fig:1}B}) plots ASD averaged FNC up-triangle and CN low-triangle. Both FNC matrices exhibited similar spatial patterns, with WM-related connections primarily distributed along the diagonal with a bigger block formation and the cross-network connections, GM-related connections tightly concentrated along the diagonal, and WM-GM connections across most of networks. Group differences were assessed using GLMs, with age and sex included as covariates. The lower triangle of the FNC difference matrix displays the estimated ASD-CN $\beta$ coefficients for all 20,301 connections, whereas the upper triangle highlights only connections that survived FDR correction at $q<0.05$ (\textbf{Fig.\ref{fig:1}C}).

\subsection{Hypo- and Hyperconnectivity Patterns in WM+GM FNC}
Based on the group-difference $\beta$ maps, we identified two distinct patterns of altered WM+GM FNC in ABIDE ASD (\textbf{Fig.\ref{fig:1}D}). Negative $\beta$ values indicated hypoconnectivity in ASD relative to CN, accounting for 17.29\% of the identified alterations, whereas positive $\beta$ values indicated hyperconnectivity, accounting for 16.56\%. The hypoconnectivity pattern was primarily associated with GM, with the GM:Paralimbic (PL) network serving as a central hub. Specifically, reduced connectivity was observed between the GM:PL network and higher-order cognitive networks, including the default mode network (DMN), salience network (SN), and insular temporal network (IT). In contrast, the hyperconnectivity pattern predominantly involved cross-tissue WM-GM connections, with the WM:Frontal (FR) network serving as a central hub. Increased connectivity was observed between the WM:FR network and higher-order cognitive networks, particularly the DMN and SN. Together, these findings highlight the involvement of higher-order cognitive networks in ASD and demonstrate the value of examining WM-GM functional interactions to characterize distinct patterns of altered brain network organization.

\subsection{Expression of ASD-Associated FNC Signatures in Infancy}
BRSs derived from WM+GM functional connectivity showed higher mean scores in ASD than CN participants and in HL than LL infants (\textbf{Fig.\ref{fig:1}E}). In ABIDE, the mean BRS was $-0.308\pm0.701$ for CN individuals and $0.130\pm0.784$ for individuals with ASD. A similar pattern was observed in infants, with mean BRSs of $-0.321\pm0.480$ for the LL group and $0.235\pm0.523$ for the HL group. These findings indicate greater mean expression of ASD-associated FNC patterns in HL infants; their relevance to individual developmental outcomes requires longitudinal follow-up.

\section{Future Work}
\label{sec:fw}
In the future, we will pursue two parallel directions of research. First, we will investigate alterations in WM+GM functional connectivity in individuals with ASD, examining their associations with behavioral measures and clinical symptom severity. This work will help us better understand the functional neural correlates of individual differences in behavioral and clinical profiles in ASD. Second, we will conduct a deeper investigation into how these neuroimaging patterns relate to developmental trajectories and their associations with infant behavioral measures, including eye-tracking indices and other developmental assessments.

\section{Acknowledgments}
The authors declare that there are no conflicts of interest related to this research. This work was supported by National Science Foundation grant 2112455, and National Institutes of Health grants R01MH123610, R01MH130595 and R01MH119251.

\section{Compliance with Ethical Standards}
This study used human subject data in accordance with the respective data-use licenses, which document ethical approval.

\bibliographystyle{IEEEbib}
\bibliography{references} 

\end{document}